\documentclass[twocolumn,showpacs,showkeys,preprintnumbers]{revtex4}
\usepackage{graphicx}
\usepackage{subfigure}
\usepackage{lscape}
\usepackage{dcolumn}
\usepackage{bm}
\usepackage{color}

\usepackage{amsmath} 
\usepackage{amssymb} 
\usepackage{multirow}
\usepackage{epstopdf}

\begin{document}
\title{Search for UHE gamma photons in cosmic rays by studying the geomagnetic influence on air-shower muons}
\author{Animesh Basak$^{1}$ \thanks{Email address: ab.astrophysics@rediffmail.com}, Meghamani Haldar$^{1}$ \thanks{Email address: haldarmeghamani@gmail.com}, Kishor Chaudhury$^{1,2}$ \thanks{Email address: chaudhurykishor@gmail.com} and Rajat K. Dey$^{1}$\email{Email address: rkdey2007phy@rediffmail.com}}
\affiliation{$^{1}$Department of Physics, University of North Bengal, Siliguri, WB, India 734013. \\
$^{2}$Department of Physics, Alipurduar University, Alipurduar, WB, India 736122.}

\begin{abstract}
A major challenge in ground-based ultra-high-energy gamma-ray observations remains in discriminating sporadic gamma-ray signals from a huge background of cosmic-ray events. To achieve good discrimination power of gamma rays against protons, an unconventional approach is presented that exploits the geomagnetic influence on air shower muons in Monte Carlo shower events. A recognizable polar asymmetry in the positive and negative muon distributions is noticed after transforming the impact coordinates of each muon from the observer plane onto the shower front plane, plus a judicious ignoring of the minimal attenuation of muons in the space between the two planes. A couple of observables, $d_{\text{max}}$ and $N_{\mu; \text{IQS}}^{\text{tr.}}$, can be extracted from the formation of two densely populated positive and negative muon zones in each shower. Both observables demonstrate excellent gamma/proton discrimination power, with efficiencies of $93.2\%$ and $98.6\%$. This analysis permitted the usage of muon detectors just covering a small area of $\simeq{1.4}\%$ of a square-km array for ultra-high-energy gamma-ray observations. It is expected that the present and future gamma-ray observatories will implement the current approach.  
  
\end{abstract}

\keywords{cosmic rays, gamma rays, extensive air showers, geomagnetic effect, simulations}
\maketitle

\section{Introduction}
One of the primary goals of any cosmic-ray air-shower experiment is to identify the nature of each cosmic-ray primary particle that enters the atmosphere as precisely and efficiently as possible. On the other hand, the selection of extensive air showers (EAS) initiated by highly energetic gamma rays ($\gamma$-rays) from the huge background of EASs that are triggered by charged hadrons, mostly by protons (p), is one of the major challenges of these experiments [1]. For illustration, the ratio between the fluxes of near isotropic $\gamma$-rays and cosmic rays (CR) has been reported to be $1:10^{5}$ at $10^{14}$~eV [2]. These limiting flux ratios between $\gamma$-rays and protons measured by some early observations were $1:10^{3}$, $1:3\times{10^{2}}$, $1:{10^{2}}$ and $1:10$ at $10^{14}$, $10^{15}$,  $10^{16}$ and $\simeq{10^{17-19}}$~eV energies [3-4]. These limits indicate that the proportion of $\gamma$-rays in background-charged CRs is the highest in $0.1-10$~EeV energy region. Moreover, only the flux of these $\gamma$-rays with energy $\gtrsim{0.1}$~EeV originated within our galaxy can survive en route to Earth owing to their strong attenuation over intergalactic distances. $\gamma$-rays are attenuated in background photon fields \textit{viz}. the cosmic microwave backround (CMB), the interstellar radiation field (ISRF) and the extragalactic bacground light (EBL) via photon-photon pair production ($\gamma\gamma \rightarrow e^{+}e^{-}$) [5-7]. One can barely detect PeV $\gamma$-rays of extragalactic origin because of their large opacity ($>1$) when the $\gamma$-ray source resides at a distance just beyond $\simeq{8.5}$~kpc from the Solar System [8]. 

The LHASSO experiment detected $534$ astrophysical $\gamma$-ray photons up to the highest energies of $1.4$~PeV which otherwise favoured the existence of PeVatron accelerators in our galaxy [9]. Recently, the HAWC collaboration reported the detection of nearly $100$ $\gamma$-ray events with energies $> 0.1$~PeV from the Galactic center (GC) region,  analysing their $7$~year of data, indicating a presence of a proton PeVatron at the GC [10]. On the other hand, $\gamma$-rays with energies $>0.1$~PeV from leptonic sources fall in the deep Klein-Nishina regime, and hence they are mostly suppressed. 

The KM3NeT experiment observed the most energetic neutrino event called KM3-230213A on February $13$, 2023 [11]. The origin of this ultra-high-energy (UHE) neutrino event of energy $0.22$~EeV has not been resolved to date. Several possible explanations for as-yet-unexplained amazing neutrino event KM3-230213A have already been asserted which include galactic, extragalatic, and cosmogenic sources with standard astrophysical explanations [12]. Some unorthodox and intriguing possibilities are also prescribed that KM3-230213A originated from the energetic Hawking radiation of exploding primordial black holes (PBHs) [13] and the decay of a super-heavy dark matter (SHDM) particle [14]. 

In the present work, we reckon that there should be a corresponding energetic $\gamma$-ray photon with energy $E > 0.1$~EeV from some these above mentioned possible sources/environments of the event KM3-230213A and if so, then how can such an UHE $\gamma$-ray photon be observed. If KM3-230213A was a cosmogenic neutrino which was produced due to the interactions of UHE CR protons with EBL and CMBR [15], then there is a plausibility of $\pi^{0}$ production from the decay of $\Delta^{+}$, which, in turn, decays into two $\gamma$-ray photons with energy $>0.1$~EeV [16]. If these $\gamma$-rays could have generated within $\leq{1-2}$~Mpc [17], there is a chance to evade further interactions and reach the Earth, preserving their energy. Recently in a different work [14], authors considered a simple scenario to investigate the possibility that the observed UHE neutrino event might have originated from the decay of SHDM. The SHDM decays into neutrino and standard model Higgs particles through a specific decay channel. Two UHE $\gamma$-ray photons can be produced from the subsequent decay of the Higgs boson with a small branching fraction, $\mathcal{O}{(10^{-3})}$ [14]. These SHDM particles should be present within $1$~Mpc from the Earth as because the genetated $\gamma$-rays with energy $>0.1$~EeV from them would possess an attenuation length between $\sim{1}$~kpc to $1$~Mpc [18]. The most favoured origin of the SHDM might be the GC [19] and some galactic high-mass X-ray binaries along with two nearby extragalctic targets such as Large Magellanic Cloud and the inner core region of Centaurus A.   

Thus, observations of $\gamma$-rays with energies $>0.1$~EeV, if any, may offer an unparalleled opportunity to probe high-energy processes in hitherto unspecified extreme astrophysical environments of the Milky Way, in particular. As stated above, $\gamma$-ray photons with energies $\gtrsim{0.1}$~EeV, are expected to be originated from $\pi^{0}$ decays, indicating the presence of hadronic CRs (usually from proton to iron) with energies more than $>1$~EeV. On the other hand, the observation of $\gamma$-ray photons even of extreme high energies, $1 - 100$~EeV and above, if any, in the Milky Way, would compel the unfolding of either some new physics or some new particle acceleration [19]. In this regard, the Pierre Auger Observatory (PAO) [20] and the Telescope Array (TA) [21] experiments have continuously been hunting for the detection of UHE $\gamma$-ray photons from the GC region of the Milky Way in particular. A consensus is that detecting a sub-dominant proportion of UHE $\gamma$-ray photons could be a smoking gun for SHDM particles [20]. 

The detection of the highly energetic $\gamma$-rays and hadronic CRs is achieved by the measurement of the longitudinal development of the EASs initiated by these particles via their collisions with other nuclei in the atmosphere [22] or by measuring the lateral density distribution of the EAS particles that hit the ground array detectors [23-24]. The selection of $\gamma$-ray showers with decent efficiency from huge charged CR-initiated showers, usually dominated by protons, is one of the challenging issues for EAS experiments employing single or hybrid detection methods [25]. Generally, a $\gamma$-ray-initiated EAS is considered to be characterized by poor muon or muon-depleted content. To select a few TeV to the PeV energy $\gamma$-ray shower based on such a criterion, an EAS array needs to have muon detectors with very good shielding framework spread across a relatively large area [9]. Economically, this is quite challenging, and such a facility is rarely attainable. At the same time, one should also remember that the intended EAS observable must have a practicable and reasonable background rejection or $\gamma$-ray selection efficiency [26].

In this paper, we have carried out an investigation into the discriminating of $\gamma$-ray showers with energy $>0.1$~EeV from the large background of proton-initiated showers, exploiting the CORSIKA Monte Carlo simulation code for CR air showers generation [27]. In recent past, we exploited a unique EAS observable named as {\lq maximum transverse muon barycenter separation (MTMBS; $d_{\text{max}}$)\rq} for studying the CR mass composition through simulations [28-29]. Additionally, those works described the need for a potential experimental layout that includes an EAS array of scintillation detectors and a pair of muon detector units. For a group of showers with some mean zenith and azimuth angles, the muon detector units were required to be planted, covering only two selected diametrically opposed regions across the shower core, rather than a distribution of muon detectors spread over the array, like LHAASO [9], for the CR mass composition study. The two muon detecting units can be rotated to be pointed in a certain direction to measure the incoming positive (${\mu}^{+}$)and negative muons (${\mu}^{-}$) corresponding to different mean zenith and azimuth angles of $\gamma$-ray and CR-intiated showers. In those works [28-29], it was specified that the two muon units would rotate on two circular tracks positioned contrariwise between mean core distances $R_{c}\simeq{60}$~m and $R_{c}\simeq{90}$~m (i.e. $\Delta{R_{c}}\in{30}$~m) respectively with $10^{\circ}$ or $15^{\circ}$ polar angle bin (i.e. $\Delta{\beta_{s}}\in{10^{\circ}}$ or $15^{\circ}$) made by two diagonal lines passing through the shower core. This work also considers the sum of those ${\mu}^{+}$ and ${\mu}^{-}$ that contributed the $d_{\text{max}}$ value of a shower, as some other gamma/proton (${\gamma}/p$) discriminant observable, called the truncated muon size, $N_{\mu; \text{IQS}}^{\text{tr.}}$. 

Section II describes details of the simulation sets and briefly their data analysis method for showers initiated by protons and $\gamma$-rays used in this work. The ${\gamma}/p$ discriminating observable, $d_{max}$, derived from a more primitive variable, the transverse muon barycenter separation (TMBS;$d_{s}$), is revisited in section III. The same section also discusses the possibility of a realistic muon estimator, called the truncated muon size, $N_{\mu; \text{IQS}}^{tr.}$, as another ${\gamma}/p$ discriminating observable along with the $d_{max}$. In section IV, the results with the efficiency of these proposed observables for selecting $\gamma$-ray events from a large background of proton events is described. Finally, section V includes a discussion briefly about possible implications of the method as well as its obtained results for the design of air shower array equipped with electron and muon detectors.

\section{Simulated data sets and their analysis procedure}
To carry out the investigation described in the introduction, we generated about $50$ events each for proton and gamma-ray showers at an energy, $E=1$~EeV using CORSIKA Monte Carlo (MC) package of version $7.7401$ [27]. The shower events were simulated at the KASCADE experiment site ($110$~m asl, $49.1^{\circ}$~N, $8.4^{\circ}$~E) [30] corresponding to the combination of EPOS-LHC [31]  and UrQMD [32] as hadronic models for high- and low-energy hadron-hadron interactions, respectively. The zenith angle ($\Theta$) was set at a fixed value, $60^{\circ}$ while the azimuthal angle ($\Phi$) of showers was always directed from the north-direction on the CORSIKA plane, i.e., $0^{\circ}$ (because of limited computing resources). At the same geographical conditions, a simulated shower sample with $50$ p-initiated and equal number of $\gamma$-initiated showers was simulated by setting the geomagnetic field (GF), $B\approx 0$ in the CORSIKA input file. Additionally, we have generated about $450$ MC showers with the $3.5:2$ ratio between $p$- and $\gamma$-initiated showers in the energy range from $0.1$ to $5$~EeV to prepare a mixed shower sample. On the observation level, the kinetic energy thresholds for the registered electrons and muons were set respectively at $0.003$~GeV and $0.3$~GeV, in the simulation.

As a first step of the shower data analysis, we have transformed each position of hit by individual muon from the observation/ground plane ($x_{g},y_{g}$ or $r_{g},\beta_{g}$) onto the shower front plane ($x_{s},y_{s}$ or $r_{s},\beta_{s}$) by employing a projection method [28-29], as given below.

\begin{equation}
r_{s}=r_{g}\sqrt{1-sin^{2}{\Theta}cos^{2}{(\beta_{g}}-\Phi)}
\end{equation}  
\begin{equation}
x_{s}=x_{g}cos{\Theta}=r_{g}cos{(\beta_{g}-\Phi)}cos\Theta
\end{equation}  
\begin{equation}
y_{s}=y_{g}=r_{g}sin{(\beta_{g}-\Phi)}
\end{equation}  

Additionally, it was shown in our previous data analysis works [28-29] that attenuation of muons in the region between the observation and shower planes was found to be unimportant. As a result, it might be disregarded in this circumstance. To make things more practical relative to muon energy thresholds and the muon detector sizes for an appreciable effect of the GF on EAS muons, we have adopted the same cuts respectively on the annular region ($60 - 90$~m) and the momenta ($p_{\mu}=10^2-10^3$~GeV/c) of ${\mu}^{+}$s and ${\mu}^{-}$s in line with our previous works in [28-29]. We have also maintained here the same $15^{\circ}$ polar angle bin by combining two diagonal lines passing through the shower core as before. 

\section{A brief theoretical description of the observables}
It is a well-known fact that the EAS muons, after their production from pions and kaons, could move for a longer duration than electrons in air because of their negligible interaction with the air nuclei, and hence muons experience the GF greatly (see, for instance, [28-29] and references therein). It was also noticed in those earlier simulation studies that relatively high-momentum muons in the EASs produced by highly inclined UHE CRs would receive much influence from the GF. As a result, the trajectories of ${\mu}^{+}$ and ${\mu}^{-}$ begin to follow some curved paths thereby generating a overall linear separation between ${\mu}^{+}$ and ${\mu}^{-}$ particles, which we designated as the TMBS [28-30,33]. The azimuthal distribution of these muons undergoes an azimuthal asymmetry, caused principally by the GF and the geometric effects, and very negligibly from the attenuation of muons [28-29,34]. After ignoring the attenuation effect for muons and then just to retain the GF effect only on them, one can remove the geometric effect on muons by a projection method as followed in the previous works [28-29]. 

Based on the projection method [28-29], the modified position of a muon in terms of their cartesian/polar coordinates in the shower front plane can be obtained from the simulated values in the ground/observation plane, and is described by ($x_{s},y_{s}$~or~$r_{s},\beta_{s}$). According to our data analysis cuts, we will consider the ${\mu}^{+}$ and ${\mu}^{-}$ particles possessing momenta $p_{\mu}=10^{2}-10^{3}$~GeV/c within an annular region between two circles with a radial core distance bin $\Delta{R_{c}}=60-90 $~m~or~$\in{30}$~m. Moreover, the annular region under consideration would be earmarked by a pair of diagonally aligned straight lines passing through the EAS core oppositely, making a central angle $\sim 15^{\circ}$ about an arbitrary polar angle $\beta_{s}$, and such a region was called {\lq{internal quadrant sector (IQS)}\rq} in [28-29]. Now drawing information of random points (hits by $\mu^{+}$ or $\mu^{-}$) from a given IQS at a particular polar position ($\beta_{s}$) of a shower in the EAS front plane, we can calculate their average position. If we have a bivariate kind of ($x,y$) distribution of $\mu^{+}$ (say), we can define the average of this distribution as
      
\begin{equation}
(\bar{X}_{\mu^{+}},\bar{Y}_{\mu^{+}}) = (\frac{\sum_{i=1}^{n} x_{\mu^{+},i}}{n}, \frac{\sum_{i=1}^{n} y_{\mu^{+},i}}{n}).
\end{equation}
Similarly, for $\mu^{-}$ from the same IQS sector, we have,
\begin{equation}
(\bar{X}_{\mu^{-}},\bar{Y}_{\mu^{-}}) = (\frac{\sum_{i=1}^{m} x_{\mu^{-},i}}{m}, \frac{\sum_{i=1}^{m} y_{\mu^{-},i}}{m}).
\end{equation}

where $n$ and $m$ indicate the total counts of $\mu^{+}$ and $\mu^{-}$ of an EAS under consideration.
The TMBS, $d_{s}^{k}$ (say, for the $k$-th shower) is estimated in each shower event as,
\begin{equation}
d_{s}^{k}=\sqrt{{(\bar{X}_{\mu^{+}}-\bar{X}_{\mu^{-}})}^{2}+{(\bar{Y}_{\mu^{+}}-\bar{Y}_{\mu^{-}})}^{2}}.
\end{equation}

To estimate the mean TMBS, $d_{s}$, from the TMBSs known, event by event, $d_{s}^{k}$, corresponding to a reasonable number of shower events (say, $N$) with a fixed primary energy ($E$), zenith angle ($\Theta$), azimuthal angle ($\Phi$) and for $p/\gamma$ primaries, we  follow the method of averaging as

\begin{equation}
d_{s}= \frac{\sum_{k=1}^{N} d_{s}^{k}}{N}.
\end{equation}

We repeat the above procedure by placing the IQS into eleven different polar positions (by changing $\beta_{s}$) at a time via some rotation of the IQS layout about the EAS core to scan the corresponding position coordinates ($x_{s}$,$y_{s}$) of $\mu^{+}$ and $\mu^{-}$ [28-29] in the same conditions described, for the $p/\gamma$-initiated showers. Thus, a polar variation of the mean TMBS, $d_{s}$ for $p/\gamma$ showers can be obtained. In [28-29], a systematic variation of $d_{s}$ with $\beta_{s}$ or IQS's position was noticed and a maximum value for the TMBS, i.e. one of our observables here, MTMBS ($d_{max}$), could be attained. Our previous works suggested that $d_{max}$ bears some basic characteristic features concerning the nature of CR primaries. 

In addition to $d_{max}$, the present work uses the sum of $\mu^{+}$ and $\mu^{-}$ (i.e. equal to $n+m$) that hit the IQS region, corresponding to its  orientation which accounts for $d_{max}$ of a shower, as yet another potential ${\gamma}/p$ discriminating observable. We designate it by $N_{\mu;\text{IQS}}^{tr.}$, and is just called the truncated muon size. 

\section{Results on $\gamma/\text{p}$ discrimination} 
\subsection{Polar asymmetry of $\mu^{+}$ and $\mu^{-}$ in EAS}
We have counted the total number of $\mu^{+}$ and $\mu^{-}$ in the IQS at its $24$ different polar positions (i.e. equivalent to a complete rotation of the IQS) in the shower plane according to our selection criteria for $R_{c}$ and $p_{\mu}$ bins. The variations of the total number of $\mu^{+}$ and $\mu^{-}$ for $\gamma$- and p-initiated showers arriving from the direction with ${\Theta}=60^{\circ}$ and $\Phi=0^{\circ}$, are shown respectively in Fig. 1a and b. 

\begin{figure}[!htbp]
	\centering
	\includegraphics[trim=0.6cm 0.6cm 0.6cm 0.6cm, scale=0.8]{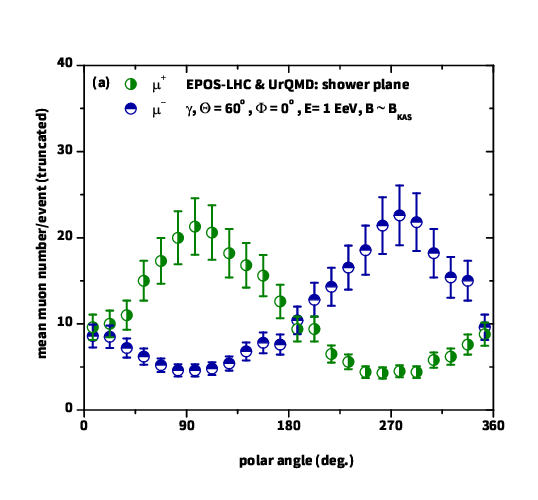}
	\includegraphics[trim=0.6cm 0.6cm 0.6cm 0.6cm, scale=0.8]{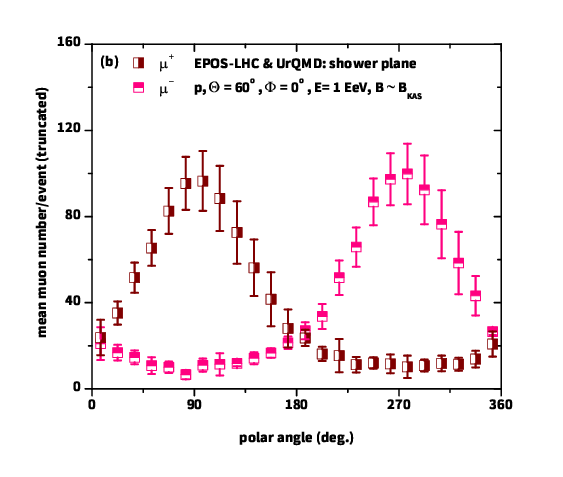}
	\caption{The mean polar variations of $\mu^{-}$ and $\mu^{+}$ for $\gamma$ events (figure a) and p events (figure b).}
\end{figure}

\begin{figure}[!htbp]
	\centering
	\includegraphics[trim=0.6cm 0.6cm 0.6cm 0.6cm, scale=0.8]{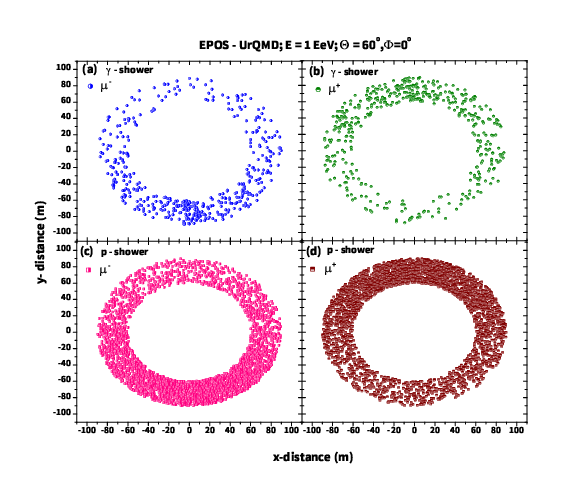}
	\caption{The mean polar variations of $\mu^{-}$ and $\mu^{+}$ for $\gamma$ events (scattered figures a \& b) and p events (scattered figures c \& d) in the $XY$-plane.}
\end{figure}
 
Irrespective of primary CR particles, coming from the geomagnetic direction (${\Theta}=60^{\circ}$ and $\Phi=0^{\circ}$), the $\mu^{+}$ and $\mu^{-}$ of their EASs experience the GF predominately around $\beta_s\simeq 90^{\circ}$ and $\simeq 270^{\circ}$, which are clearly revealed by Fig. 1a and b. For these showers, we have included scattered plots as well through Fig. 2a - d to provide a clear view of the asymmetries formed in the $\mu^{+}$ and $\mu^{-}$ distributions. 

\subsection{The discriminant observables: $d_{\text{max}}$ \& $N_{\mu;\text{IQS}}^{tr.}$}
It is evident from Fig. 1 or Fig. 2 that the TMBS, $d_{s}$ of muons, is likely to change with the mean polar position of the IQS. For the average p- and $\gamma$-initiated showers, the nature of variation of the $d_{s}$ versus $\beta_s$ plot due to a counter clockwise rotation, $-180^{\circ}$ to $180^{\circ}$, of the IQS is shown in Fig. 3. Here, $d_{s}$ takes higher values corresponding to the orientation of the IQS through $\beta_{s}\sim 90^{\circ} - 270^{\circ}$ with $\delta{\beta_{s}}\simeq{15^{\circ}}$, which are as per expectations. To understand the key role of the GF in the present $\gamma/p$ discrimination method, we have also studied the polar variation of $d_{s}$ for an equal number of p- and $\gamma$-initiated showers ($50$ events each for $p$ and $\gamma$ primaries) with same $E$, $\Phi$ and $\Theta$ as above but setting $B\simeq 0$, and are shown in the same figure. Thus, it demonstrates that the correlation of the variable $d_{s}$ with $\beta_s$ in a real situation (i.e. non-zero Earth's GF) enhances the efficiency of the observable, $d_{\text{max}}$ to discriminate $\gamma$ events from the proton event dominated sample.

\begin{figure}[!htbp]
	\centering
	\includegraphics[trim=0.6cm 0.6cm 0.6cm 0.6cm, scale=0.8]{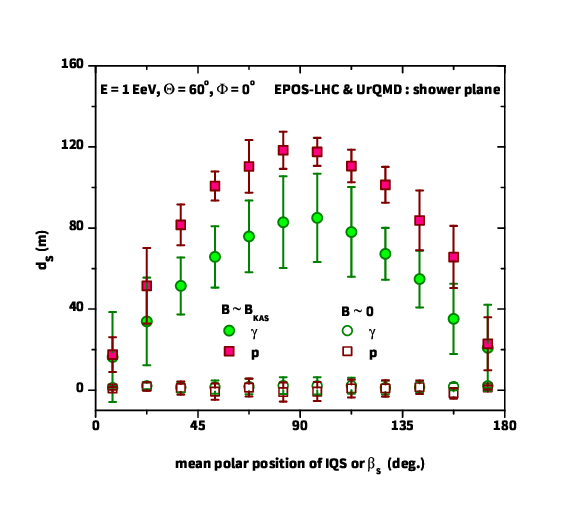}
	\caption{The mean polar variations of the TMBS ($d_{\text{s}}$) for $\gamma$ events (filled circles) and p events (filled squares) when $B\simeq{B_{\text{KAS.}}}$. The respective mean polar variations for $\gamma$ events (unfilled circles) and p events (unfilled squares) are also shown when $B\simeq{0}$.}
\end{figure}

\begin{figure}[!htbp]
	\centering
	\includegraphics[trim=0.6cm 0.6cm 0.6cm 0.6cm, scale=0.8]{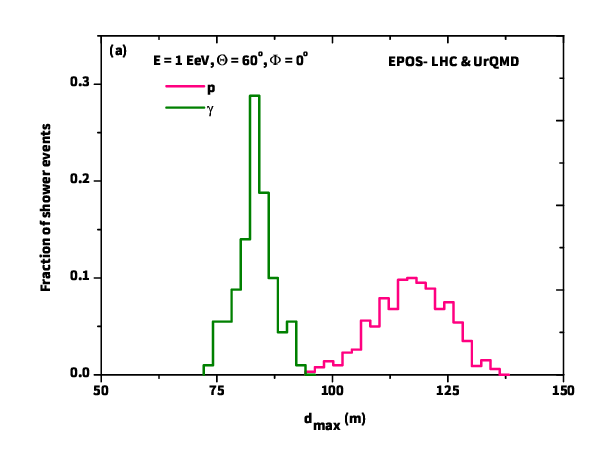}
	\includegraphics[trim=0.6cm 0.6cm 0.6cm 0.6cm, scale=0.8]{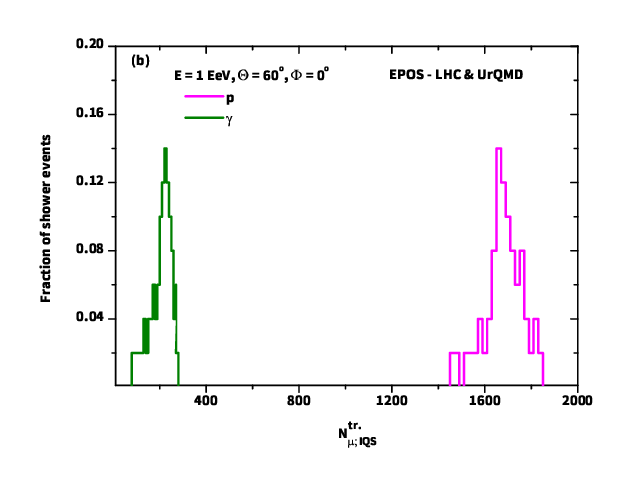}
	\caption{Distributions (figure a) of $d_{\text{max}}$ for p events (pink line) and $\gamma$ events (olive green line) at a fixed energy. On the bottom (figure b), the respective distributions of $N_{\mu;\text{IQS}}^{tr.}$ are shown.}
\end{figure}

The average $d_{s}$ profile with $B\simeq B_{\text{KAS.}}$, as shown in Fig. 3, has been obtained for each MC shower in a given sample from our data analysis. This allows one to extract the discriminant observables, $d_{\text{max}}$, and the corresponding truncated muon size, $N_{\mu;\text{IQS}}^{tr.}$ on an event-by-event basis. The $d_{\text{max}}$, as well as the $N_{\mu;\text{IQS}}^{tr.}$ distributions for these p and $\gamma$ events are shown in Fig. 4a and b. These figures indicate that the $\gamma/p$ separation power of the probing observables based on the geomagnetic spectroscopy method on air-shower muons is ensuring their potential use for future UHE ground-based gamma-ray observatories.

However, there is a scope to mimic some realistic experimental situation here if we could prepare a mixed sample with a relatively lesser number of $\gamma$ events and, check to see if it is possible to extract those $\gamma$ events out. We prepared a mixed CR sample with an energy ranging from $0.1$ to $5$~EeV while keeping the ratio of primaries at $\simeq 78\%$ p and $\simeq 22\%$ $\gamma$ events corresponding to $\Phi=0^{\circ}$ and $\Theta=60^{\circ}$. We have made use only the EPOS-UrQMD model combination as the high- and low- energy hadronic interactions. Hence, the utility in examining other hadronic interaction models as well as the $\Phi$ and $\Theta$ bins, go beyond the scope in the current article  because of the limited computational resources.

For producing some substantial results using the mixed sample, we have chosen the p and $\gamma$ events that lie in a specific energy bin $0.2-0.6$~EeV. This choice further modifies the mixture ratio with $288$ number of p events and $74$ number of $\gamma$ events. This option of the energy provides a reasonable statistics of EAS events because of the steeply falling nature of the CR flux ($\gamma$ flux) $E^{-3}$ ($E^{-2}$). The energy bin also includes the $\gamma$ photon(s) that might have associated with the KM3-230213A neutrino event or some hitherto unknown origins.

Figure 5 is an illustration of the correlation between the $d_{\text{max}}$ and the $N_{\mu;\text{IQS}}^{tr.}$ for MC p and $\gamma$ events of the  mixture in the energy range $0.2-0.6$~EeV. Although the two observables seem to unveil some indiscriminate correlation here but it still exhibits their ability to discriminate UHE $\gamma$events from the background of proton-initiated events. Accordingly, the $d_{\text{max}}$ and  $N_{\mu;\text{IQS}}^{tr.}$ distributions for MC p and $\gamma$ events of the mixture are shown independently in Fig. 6a and b. Events from the mixture have been selected choosing the respective identification (ID) codes (we set ID code 1 for $\gamma$ and 2 for p events) assigned to the analyzed showers before their mixing. Overall, the $N_{\mu;\text{IQS}}^{tr.}$ provides an excellent $\gamma/p$ discriminator (Fig. 6b), and therefore, it is combined with our prioritize $\gamma/p$ discriminator $d_{\text{max}}$ (Fig. 6a) in the present analysis.     

\subsection{$\gamma/p$ discrimination}
Finally, we put effort into selecting the proportion of $\gamma$ events out of the mixture using some selection procedure which imposes different selection cuts values on the  parameters $d_{\text{max}}$ or the $N_{\mu;\text{IQS}}^{tr.}$ within their respective two extremes. These cut values enable different acceptances or rejections of $\gamma$ and p events, respectively, in the concerned energy range. The astronomical data analysis method sets optimal cuts on selection parameters by numerical maximization of the quality factor ($Q$) [25], defined by
\begin{equation}
Q= \frac{\epsilon_{\gamma}}{\sqrt{\epsilon_{p/\text{bkg}}}}
\end{equation}
where $\epsilon_{\gamma}$ and $\epsilon_{p/{\text{bkg}}}$ represent respectively the acceptances of $\gamma$ and p events from the mixture using an arbitrary cut value of $d_{\text{max}}$ or $N_{\mu;\text{IQS}}^{tr.}$ within the two extremes. Alternatively, the rejection of p-initiated or background events can be expressed by $\xi_{p/\text{bkg}}=1-\epsilon_{p/\text{bkg}}$. Here, the value of $Q$ determines the effectiveness of the observables $d_{\text{max}}$ or $N_{\mu;\text{IQS}}^{tr.}$ in $\gamma/p$ discrimination. 

The acceptance efficiencies of $\gamma$ and p events from the mixture are defined by $\epsilon_{\gamma}=\frac{n_{\gamma{(\text{cut})}}}{n_{\gamma}}$ and $\epsilon_{p}=\frac{n_{p}{(\text{cut})}}{n_p}$, where $n_{\gamma}{(\text{cut})}$ and $n_p{(\text{cut})}$ are the number of $\gamma$ and p events that passed a specific cut value of the $d_{\text{max}}$ or $N_{\mu;\text{IQS}}^{tr.}$.  Here, the total number of $\gamma$ and p events that are present in the mixed data-set with $E$-range $0.2-0.6$~EeV, are represented by $n_{\gamma}$ ($74$ events) and $n_{p}$ ($288$ events). We have estimated the quality factor by setting different cut limits on $d_{\text{max}}$ or $N_{\mu;\text{IQS}}^{tr.}$ within the two respective extremes and are compiled in Table 1 and Table 2.  
   
\begin{figure}[!htbp]
	\centering
	\includegraphics[trim=0.6cm 0.6cm 0.6cm 0.6cm, scale=0.8]{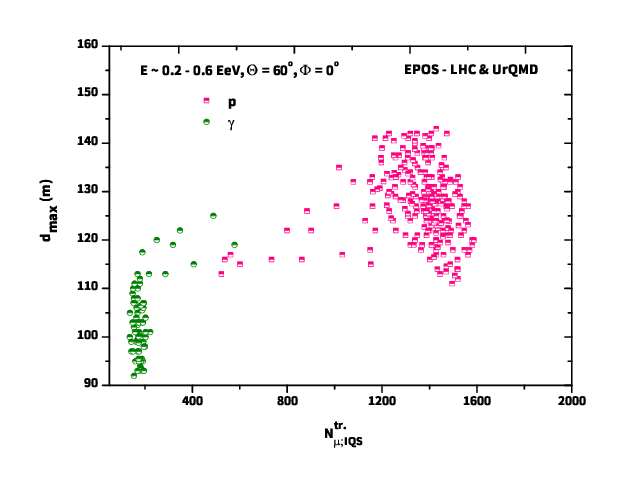}
	\caption{Scatter plots of $d_{\text{max}}$ and $N_{\mu;\text{IQS}}^{tr.}$ for $p$ (half filled pink squares) and $\gamma$ (half filled olive green circles) events segregated from the mixed sample with energy ranging from $0.2$~EeV to $0.6$~EeV.}
\end{figure}

\begin{figure}[!htbp]
	\centering
	\includegraphics[trim=0.6cm 0.6cm 0.6cm 0.6cm, scale=0.8]{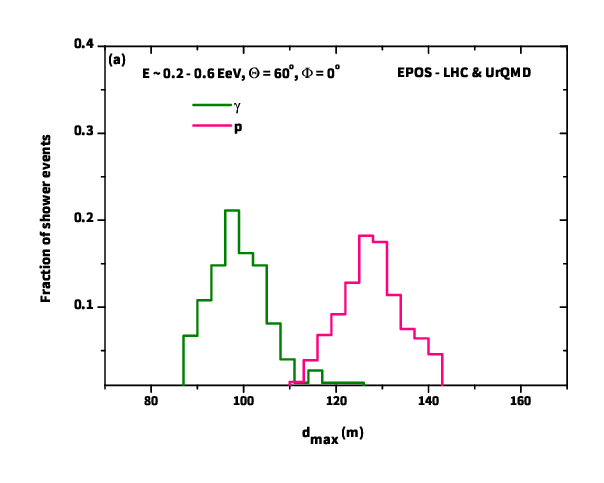}
	\includegraphics[trim=0.6cm 0.6cm 0.6cm 0.6cm, scale=0.8]{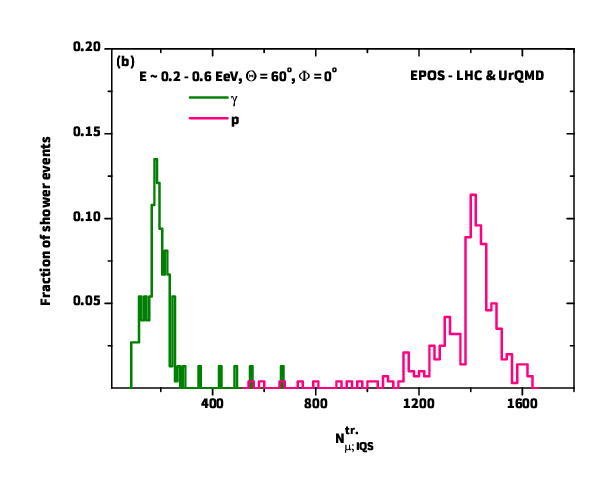}
	\caption{Distributions (figure a) of the $d_{\text{max}}$ for p events (pink line) and $\gamma$ events (olive green line) with energy ranging from $0.2$~EeV to $0.6$~EeV. On the bottom (figure b), the respective distributions of the $N_{\mu;\text{IQS}}^{tr.}$ are shown.}
\end{figure}

\begin{table}[!htbp]
	\begin{tabular}{|l|l|l|l|l|l|l|l|}
		\hline
     $d_{\text{max.}}$    &$110.0$    &$112.5$    &$115.0$    &$117.5$    &$120.0$    &$122.5$    &$125.0$   \\ \hline
    
     $\epsilon_{\gamma}$    &$0.892$    &$0.905$    &$0.932$    &$0.972$    &$0.986$    &$1.0$    &$0$ \\ \hline
    
     $\epsilon_{p/\text{bkg}}$    &$0$    &$0$    &$0.038$    &$0.063$    &$0.142$    &$0.211$    &$0.330$ \\ \hline

     $Q$    &$\infty$ (Ideal case)    &$\infty$ (Ideal case)    &$\bf{4.78}$    &$3.87$    &$2.62$    &$2.18$    &$0$ (NGS)  \\ \hline 
\end{tabular}
	\caption{The quality factor at various $d_{\text{max}}$ cuts. The highlighted $Q$ corresponds to the optimal cut value of the $d_{\text{max}}$. NSG stands for {\lq no $\gamma$ signal\rq}.}
\end{table}

Table 1 indicates that at optimal cut with $d_{\text{max}}\simeq 115$~m, $Q$ takes a value close to $4.78$ which selects the highest proportion of $\gamma$ events ($\simeq 93.2\%$ of the total $74$ events in the mixture) with the poorest proportion of p events ($\simeq 3.8\%$ of the total $288$ events in the mixture). At a cut value of $d_{\text{max}}\simeq 122.5$~m, we could even achieve selecting $100\%$ of the $\gamma$ events from the mixture, but the relative proportion of the background events became substantial ($\simeq 21.1\%$) compared to the selection worked out at optimal condition. 

\newpage

\begin{table}[!htbp]
	\begin{tabular}{|l|l|l|l|l|l|}
		\hline
     $N_{{\mu};\text{IQS}}^{\text{tr.}}$    &$600$    &$700$    &$800$    &$900$    &$1000$   \\ \hline
    
     $\epsilon_{\gamma}$    &$0.986$    &$1.00$    &$0$    &$0$    &$0$    \\ \hline
    
     $\epsilon_{p/\text{bkg}}$    &$0.007$    &$0.010$    &$0.018$    &$0.025$    &$0.036$   \\ \hline

     $Q$    &$\bf{11.8}$    &$10.0$    &$0$ (NGS)    &$0$ (NGS)    &$0$ (NGS)   \\ \hline 
\end{tabular}
	\caption{The quality factor at various $N_{\mu;\text{IQS}}^{tr.}$ cuts. The highlighted $Q$ corresponds to the optimal cut value of the $N_{\mu;\text{IQS}}^{\text{tr.}}$. NSG stands for {\lq no $\gamma$ signal\rq}.}
\end{table}

Using $N_{\mu;\text{IQS}}^{\text{tr.}}$, the other discriminant variable, we achieved a fairly better result in relation to $d_{\text{max}}$, and is shown in Table 2. Here,  about $98.6\%$ $\gamma$ showers could be separated from the mixture with only less than $\simeq 1\%$ (actually $0.7\%$) p events as background corresponding to the optimal cut on $N_{\mu;\text{IQS}}^{\text{tr.}}\simeq 600$, giving $Q\simeq 11.8$. Thus, the muon detectors having charge identifying capability (which is needed for the measurement of $d_{\text{max}}$) of a gamma-ray observatory, located only in the IQS region through $\beta_{s}\sim 90^{\circ}-270^{\circ}$ with $\Delta{\beta_s}\simeq 15^{\circ}$ corresponding to our simulation sets, ensuring a very good p/background rejection levels in the concerned energy region.             

\section{Discussion and Conclusions}

An unambiguous detection of UHE photons and their separation from a huge hadronic background continues to be a primary challenge for a gamma-ray observatory. This simulation study aims to select highly inclined $\gamma$-induced showers from those generated by protons based on the influence of the Earth's magnetic field on EAS muons that finally reach the detector level. The total number of muons ($N_{\mu}$) of a shower arriving at the underground muon detector array, is often regarded as a reliable $\gamma/\text{hadron}$ discriminator. However, the setup of such an effective measurement technique is very expensive, particularly at the UHE range (requiring a few km$^2$ array). 

Here, we first tried applying our data analysis technique to retain principally the GF effects on muons of simulated $\gamma$ and p showers. Then, we identified two opposite circular arc sectors in a specific IQS region that account a highly contrasting polar asymmetry between $\mu^{+}$ and $\mu^{-}$ of a $\gamma/p$-induced shower (here, it is through $\beta_{s}\sim 90^{\circ}-270^{\circ}$ for showers with $\Phi=0^{\circ}$ and $\Theta=60^{\circ}$). We simply need a comparatively smaller number of muon detectors with charge identifying features (for $d_{\text{max}}$ estimation) covering both the circular arc sectors of the IQS (area is $\frac{\pi{(90^{2}-60^{2})}}{12}$=1177.5~m$^{2}$, which is nearly $0.12\%$ of a square-km array) with $\Delta{R_{c}}=30$~m, $\delta{\beta_s}=15^{\circ}$ for such a set of showers. This tactic effectively lowers spending concerning the $N_{\mu;\text{IQS}}^{\text{tr.}}$ measurement. Further, it might be feasible to use it for another set of showers (with different $\Phi$ and/or $\Theta$) if some circular rail tracks are built respectively at $R_{c}\approx 60$~m and $R_{c}\approx 90$~m that can transport the muon detector assembly from its present site at the two arc sectors of the IQS to some other site of different polar position . If there is no financial contraints, the most effective approach would require the installation of muon detectors across the entire annular region between radii of $60$~m and $90$~m. The resulting surface area of the muon detectors would be about $12$ times that of the above configuration, and it is equal to $\simeq{0.014}$~km$^{2}$, which is nearly $1.4\%$ of a square-km array. Still, the muon detector area is relatively much smaller compared to that of a few km$^2$ array at UHE energies.

The determination of muon charge signs and their impact coordinates ($x,y$) necessitates a non-magnetized muon detection system comprising shielded scintillators, which is effective in determining the charge sign of muons by measuring their lifetime within a vertically arranged stack of active (plastic scintillator sheets) and passive (aluminum plates) layers. Details regarding the potential muon detecting systems are available in previous studies [28-29,35]. A densely packed array of electron detectors in conjuction with the unique muon detectors as reviewed above would be needed to obtain the standard shower parameters, like $\Theta$, $\Phi$, core location ($x_{0},y_{0}$), and shower size or $E$ from the density and timing information of the electromagnetic component of an EAS. The meticulous estimation of the EAS shower cores is essential to the accuracy of data on impact coordinates $x,y$ of muons and other EAS observables. For more precise estimation of the shower cores, a closely spaced array of electron detectors is preferable. 

Investigating other sets of MC events with different $\Theta$, $\Phi$, or their bins lies beyond the scope of the current analysis. Examining the influence of various hadronic interaction models is also outside the scope of this article. It is noteworthy that the present method was found ill-suited to discriminate $\gamma$ showers from p events in the TeV to even a few tens of PeV energy range. The main reason is the very low number of muons registered in the specified IQS region, particularly in $\gamma$-initiated events.

We noticed that the $\gamma$ separation efficiency is reasonably better with respect to $N_{\mu;\text{IQS}}^{\text{tr.}}$ compared to the case with $d_{\text{max}}$ at optimal conditions. It is justifiable in the sense that the information related to the muon number, i.e. $N_{\mu;\text{IQS}}^{\text{tr.}}$, can be obtained in a single step (transforming each position of hit by individual muon from the ground plane to the shower front plane) from the simulation. On the other hand, $d_{\text{max}}$ estimation involves some algebraic mean, square, and square root operations using muon data from the shower front plane. These multiphase calculations may affect the $d_{\text{max}}$ estimate.

\section*{Data availability statement}
	All data that support the finding of the study are obtained using cosmic-ray Monte Carlo air shower code, CORSIKA: therefore this manuscript has no associated real data.

\end{document}